\documentclass[reqno,12pt]{amsart}
\usepackage{amsmath}
\usepackage{amsxtra}
\usepackage{amscd}
\usepackage{amsthm}
\usepackage{amsfonts}
\usepackage{amssymb}
\usepackage{eucal}
\usepackage[matrix,arrow,curve]{xy}
\usepackage{cite}
\usepackage{mathtools}
\theoremstyle{plain}
\newtheorem{Thm}[subsection]{Theorem}
\newtheorem{Cor}[subsection]{Corollary}
\newtheorem{Lem}[subsection]{Lemma}
\newtheorem{Prop}[subsection]{Proposition}
\newtheorem{Conj}[subsection]{Conjecture}

\theoremstyle{definition}
\newtheorem{Def}[subsection]{Definition}

\theoremstyle{remark}

\newtheorem{Rem}[subsection]{Remark}

\newtheorem{conj}{Conjecture}

\numberwithin{equation}{section}
\renewcommand{\rm}{\normalshape}

\newcommand{\refl}[1]{Lemma ~\ref{L:#1}}

\newcommand{\bref}[1]{\textbf{\ref{#1}}}

\newenvironment{thm}[1]%
    { \begin{Thm} \label{T:#1}  \ifShowLabels \TeXref{T:#1} \fi }%
    { \end{Thm} }

\renewcommand{\th}[1]{\begin{thm}{#1} \sl }
\renewcommand{\eth}{\end{thm} }

\newenvironment{lemma}[1]%
    { \begin{Lem} \label{L:#1}  \ifShowLabels \TeXref{L:#1} \fi }%
    { \end{Lem} }
\newcommand{\lem}[1]{\begin{lemma}{#1} \sl}
\newcommand{\elem}{\end{lemma}}

\newenvironment{propos}[1]%
    { \begin{Prop} \label{P:#1}  \ifShowLabels \TeXref{P:#1} \fi }%
    { \end{Prop} }
\newcommand{\prop}[1]{\begin{propos}{#1}\sl }
\newcommand{\eprop}{\end{propos}}

\newenvironment{corol}[1]%
    { \begin{Cor} \label{C:#1}  \ifShowLabels \TeXref{C:#1} \fi }%
    { \end{Cor} }
\newcommand{\cor}[1]{\begin{corol}{#1} \sl }
\newcommand{\ecor}{\end{corol}}

\newenvironment{defeni}[1]%
    { \begin{Def} \label{D:#1}  \ifShowLabels \TeXref{D:#1} \fi }%
    { \end{Def} }
\newcommand{\defe}[1]{\begin{defeni}{#1} \sl }
\newcommand{\edefe}{\end{defeni}}

\newenvironment{remark}[1]%
    { \begin{Rem} \label{R:#1}  \ifShowLabels \TeXref{R:#1} \fi }%
    { \end{Rem} }
\newcommand{\rem}[1]{\begin{remark}{#1}}
\newcommand{\erem}{\end{remark}}

\newenvironment{conjec}[1]%
    { \begin{Conj} \label{Co:#1}  \ifShowLabels \TeXref{Co:#1} \fi }%
    { \end{Conj} }
\renewcommand{\conj}[1]{\begin{conjec}{#1} \sl }
\newcommand{\econj}{\end{conjec}}

\newcommand{\eq}[1]%
    { \ifShowLabels \TeXref{E:#1} \fi
       \begin{equation} \label{E:#1} }
\newcommand{\eeq}{ \end{equation} }

\newcommand{\prf}{ \begin{proof} }
\newcommand{\epr}{ \end{proof} }

\def\be{\begin{equation}}
\def\ee{\end{equation}}
\def\ba{\begin{array}}
\def\ea{\end{array}}

\newif\ifShowLabels
\ShowLabelstrue
\newdimen\theight
\def\TeXref#1{%
    \leavevmode\vadjust{\setbox0=\hbox{{\tt
        \quad\quad  {\small \rm #1}}}%
    \theight=\ht0
    \advance\theight by \lineskip
    \kern -\theight \vbox to
    \theight{\rightline{\rlap{\box0}}%
    \vss}%
    }}%
\ShowLabelsfalse% comment this out if labels should be printed

\numberwithin{equation}{section} \makeatletter
\@addtoreset{equation}{section}

\makeatletter
\g@addto@macro{\endabstract}{\@setabstract}
\newcommand{\authorfootnotes}{\renewcommand\thefootnote{\@fnsymbol\c@footnote}}%
\makeatother

\begin{document}

\begin{flushright}
FIAN-TD-2014-10 \\
\end{flushright}

\bigskip
\bigskip
\begin{center} 
{ \large\textbf{ Frobenius manifolds, Integrable Hierarchies
 \\ \vspace{3mm} and Minimal Liouville Gravity }}
 \par
 \vspace{.8cm}

  \normalsize
  \authorfootnotes
A.~A. Belavin\footnote{belavin@itp.ac.ru}\textsuperscript{1,3,4} and
V.~A. Belavin\footnote{belavin@lpi.ru}\textsuperscript{2,3}\par \bigskip

\begin{tabular}{ll}
$\qquad^{1}$~\parbox[t]{0.9\textwidth}{\normalsize\raggedright
\small{L.~D. Landau Institute for Theoretical Physics, 142432 Chernogolovka, Russia}}\\
$\qquad ^{2}$~\parbox[t]{0.9\textwidth}{\normalsize\raggedright
\small{P.~N. Lebedev Physical Institute, 119991 Moscow, Russia}}\\
$\qquad^{3}$~\parbox[t]{0.9\textwidth}{\normalsize\raggedright
\small{Institute for Information Transmission Problems,  127994 Moscow, Russia}}\\
$\qquad^{4}$~\parbox[t]{0.9\textwidth}{\normalsize\raggedright
\small{Moscow Institute of Physics and Technology, 141700 Dolgoprudny, Russia}}
\end{tabular}\\
%  \today
\end{center}
\vspace{.8cm}
\begin{abstract}
We use the connection between the Frobrenius manifold and the Douglas string equation
to further investigate Minimal Liouville gravity. We search a solution
of the Douglas string equation and simultaneously a proper transformation from the KdV to the Liouville
frame which ensure the fulfilment of the conformal and fusion selection rules.
We find that the desired solution of the string equation has explicit and simple form in the flat 
coordinates on the  Frobenious manifold in the general case of (p,q) Minimal Liouville gravity. 
\end{abstract}
 \bigskip

\section{Introduction}
\label{sec:Introduction}

The purpose of this paper is to further study Minimal Liouville Gravity (MLG)\cite{Polyakov:1981rd}
using an approach based on the  Douglas string equation \cite{Douglas:1989dd}. 
This study is a continuation of earlier  works \cite{Moore:1991ir,
Belavin:2008kv,Belavin:2013,VBelavin:2014fs}.

The Liouville Gravity  represents a consistent example of the noncritical String theory.
In the initial continuous approach  the Liouville Gravity is formulated as a BRST invariant theory composed of 
the matter sector, the Liouville theory and the ghosts system. 
MLG represents the theory, where
the matter sector is taken to be a $(p,q)$ Minimal Model of CFT \cite{Belavin:1984vu}.
The main problem of MLG is to calculate correlation functions of BRST invariant observables, 
which are given by integrals over moduli of  Riemannian surfaces. Usually they are called the correlation numbers.
Numerous examples show that the solution of the problem is quite nontrivial within the framework of the 
continuous approach.  

An alternative approach to MLG has grown up from the idea of triangulations
of two-dimensional surfaces realized in terms  of Matrix Models \cite{Kazakov:1985ea, Kazakov:1986hu, Kazakov:1989bc, Staudacher:1989fy, Brezin:1990rb, Douglas:1989ve, Gross:1989vs}. 
One of the most important points  of the approach is the String equation which was derived by Douglas  \cite{Douglas:1989dd} in Matrix Models approach to two dimensional gravity.
The subject of the String equation is  the generating function of the correlation numbers which depends on the parameters of the problem (the  so called KdV times).
In our work, following \cite{Belavin:2008kv,Belavin:2013,VBelavin:2014fs}, we will conjecture  that the Douglas equation is applicable to  the Minimal Liouville gravity as  well as to  Matrix Models of
2D gravity.

 This   conjecture requires  the following two questions to be answered:
 how to choose the desired solution of the Douglas string equation and an appropriate form of the
so called resonance transformation \cite{Moore:1991ir} from the KdV times to the Liouville coupling constants. 
Once these two questions are answered, the generating function of the correlation functions
in MLG  is given explicitly as an integrated one-form defined uniquely for each $(p,q)$ MLG model
and coincides with a special choice of the tau-function of the dispersionless limit \cite{Krichever:1992sw,Dubrovin:1992dz}
of the generalized KdV hierarchy. 

In this paper, using the connection \cite{Belavin:2013} of the approach  to MLG \cite{Belavin:2008kv}
based on the String equation  with the Frobenius manifold structure,  
we  find the necessary solution of the String equation.
We also show that this solution together with the suitable choosen resonance transformation
lead to the results which are consistent with the main requirements of $(p,q)$ models of MLG 
(the so called  selection rules).
It is remarkable  that the needed  solution of the  Douglas equation has a very simple form in the flat 
coordinates on the  Frobenious manifold in the general case of (p,q) Minimal Liouville
 as well as it has been found recently in the case of Unitary  models of MLG \cite {VBelavin:2014fs}.

The paper is organized as follows.
In Section \bref{sec:FrobMan} we recall briefly the  notion of the Frobenius manifold and discuss its
basic properties. 
In Section \bref{sec:MainExm} we discuss the Frobenius manifold that appears 
in the context of Minimal Liouville Gravity. 
 Section \bref{sec:FM_IH} is devoted to the connection
between the Frobenius manifold structures, Integrable structures and the Douglas string equation.
In Section \bref{sec:resonance} we focus on the $(p,q)$ models of MLG and discuss the problem
of the resonance transformations.  
The idea of the  approach based on the String equation  to $(p,q)$ MLG is formulated 
in Section \bref{sec:SolutionPlan}. 
The  appropriate solution of the Douglas string equation is discussed in Section \bref{sec:AppSol}.
The rest of the paper is devoted to the analysis of the correlation functions. 
We show that   the special choice of the solution of the String equation together with the resonance
transformation encoded in terms of Jacobi polynomials ensure  fulfilling  the necessary selection rules
for the correlation numbers in  $(p,q)$ MLG.

\section{Frobenius manifolds}
\label{sec:FrobMan}
In this  and two next sections  we give  the definition and a short review of  main properties 
of the Frobenius manifolds needed for our  purposes. Here we  follow the paper by B.Dubrovin  \cite{Dubrovin:1992dz}, see also  \cite{Belavin:2013}.

 By definition  a commutative associative algebra $A$ with unity  equipped
with a nondegenerate invariant bilinear form $(\, , ) $  is called Frobenius algebra.
The invariance of the bilinear form means that for any three vectors $a,b,c$ in $A$:
\begin{equation}\label{inva}
(a\cdot b, c)=(a, b\cdot c).
\end{equation}

Let $M$ be $n$-dimensional manifold with a flat metric $\eta_{\alpha\beta} dv^ {\alpha}  dv^ {\beta}$
 which is constant in the flat  coordinates $v^{\alpha}$.

We introduce in the tangent space $T_{\bf v} M$ the structure of the Frobenius algebra 
by the following identification of the bases 
\be 
\frac{\partial }{\partial v^{\alpha}}\rightarrow e_{\alpha},\qquad
\ee
Thus,  we can multiply tangent vectors at any point of  $M$
\be
 e_{\alpha}e_{\beta}=C_{\alpha\beta}^{\gamma} e_{\gamma}.
\ee
The structure constants $C_{\alpha\beta}^{\gamma} $ may depend on  $v^{\alpha} $.
Such manifold $ M$ can be called quasi-Frobenius manifold.

\defe{FM}
The manifold $M$ is called Frobenius manifold if these two  structures
are adjusted with each other in such a way  that

(1)  the invariant bilinear form  $(\frac{\partial }{\partial v^{\alpha}},\frac{\partial }{\partial v^{\beta}} ) $  
is identical to  the  metric $\eta_{\alpha\beta} $ ;

(2) the structure  of the Frobenius algebra  at each point of $M$ and  the metric on $M$  are constrainted 
by the following relation 
\be\label{FrobM}
\nabla_{\rho} C_{\alpha\beta\gamma}=\nabla_{\alpha} C_{\rho\beta\gamma}.
\ee
\edefe
The last requirement is equivalent to the requirement that there exists a function $F$ on $M$  which is connected 
with the structure constants of the Frobenius algebra as
\be
 C_{\alpha\beta\gamma}=\frac{\partial^3 F}{\partial v^\alpha\partial v^\beta\partial v^\gamma},
\ee
where
\be
 C_{\alpha\beta\gamma}=\eta_{\alpha\rho} C_{\beta\gamma}^{\rho}.
\ee
Function $F$ is called Frobenius potential. The consistency of this property with the 
associativity of the Frobenius algebra is known as  WDVV condition \cite{Dijkgraaf:1990dj}
\be\label{WDVV}
\frac{\partial^3 F}{\partial v^\alpha\partial v^\beta\partial v^\rho}
\, \eta^{\rho\lambda}\,  \frac{\partial^3 F}{\partial v^\lambda\partial v^\mu\partial v^\nu}=
\frac{\partial^3 F}{\partial v^\nu\partial v^\beta\partial v^\rho}\, \eta^{\rho\lambda} \,
\frac{\partial^3 F}{\partial v^\lambda\partial v^\mu\partial v^\alpha}.
\ee 
The following statement  \cite {Dubrovin:1992dz} follows from these properties of the  Frobenius manifold $M$.
There exist an one-parametric flat deformation $\widetilde{\nabla}_{\alpha}$ of the connection $\nabla_{\alpha}$
\be
\widetilde{\nabla}_{\alpha}x^{\gamma}=\nabla_{\alpha} x^{\gamma}-z C_{\alpha\beta}^{\gamma} x^{\beta},
\ee
or, equivalently,
\be\label{nablacomm}
 [\widetilde{\nabla}_{\alpha}(z),\widetilde{\nabla}_{\beta}(z)]=0.
\ee
The proof is based on the associativity of the Frobenius algebra and the equation  \eqref{FrobM}.
As a consequence of \eqref{nablacomm},  there exist $n$ linear independent solutions 
\be 
\theta^{\alpha}(v,z)=\sum_{k=0}^{\infty} \theta_k^{\alpha}(v) z^k,
\ee
of the equation $\widetilde{\nabla}_{\alpha} d \theta^{\lambda}(v,z)=0$, which is equivalent to
\be
\frac{\partial^2 \theta^{\lambda}}{\partial v^{\alpha}\partial v^{\beta}}(v,z)=
z C_{\alpha\beta}^{\gamma}\frac{\partial\theta^{\lambda}}{\partial v^{\gamma}}(v,z),
\ee
or
\be\label{recur}
\frac{\partial^2 \theta^{\lambda}_{k+1}}{\partial v^{\alpha}\partial v^{\beta}}(v)=
C_{\alpha\beta}^{\gamma}\frac{\partial\theta^{\lambda}_k}{\partial v^{\gamma}}(v).
\ee
The functions $ \theta^{\alpha}(v,z)$ can be considered as the flat coordinates of the deformed 
connection $ \widetilde{\nabla}_{\alpha}(z) $.
We choose $\theta^{\lambda}(v,z)$ so that $\theta^{\lambda}(v,0)=\theta^{\lambda}_0(v)=v^{\lambda}$.
From \eqref{recur} it follows, that
\be
\nabla(\nabla \theta^{\alpha}(v,z_1),\nabla \theta^{\beta}(v,z_2))=
(z_1+z_2)\nabla \theta^{\alpha}(v,z_1)\cdot\nabla \theta^{\beta}(v,z_2),
\ee
and, hence, the scalar product $(\nabla \theta^{\alpha}(v,z),\nabla \theta^{\beta}(v,-z))=Const(z)$ 
does not depend on $v^{\alpha}$. For $z=0$ we find $Const(0)=\eta^{\alpha\beta}$.
Equation \eqref{recur} is invariant with respect to the transformation
\be
\theta^{\mu}(v,z)\rightarrow A_\nu^\mu (z) \theta^{\nu}(v,z),
\ee
where $A_\nu^\mu (0)=\delta_\nu^\mu$.
Using these transformations one can fix the normalization in such a way that
\be
(\nabla \theta^{\alpha}(v,z),\nabla \theta^{\beta}(v,-z))=\eta^{\alpha\beta}.
\ee

\section{Main example: Frobenius manifold of $A_{q-1}$-type}
\label{sec:MainExm}

Our main example is $A_{q-1}$ Frobenius manifold \cite {Dijkgraaf:1990dj}. Let $Q(y)$ be a polynomial of $y$ 
\be
Q(y)=y^{q}+u_1 y^{q-2}+...+u_{q-1},
\ee 
and $\{u_{\alpha}\}$ represent some coordinates on $M$. We call
$\{u_\alpha\}$ the canonical coordinates.
\defe{AnFM} $A_{q-1}$ Frobenius algebra is the space of polynomials modulo polynomial $\frac{d Q}{d y}$:
\be
A_{q-1} (u)=\mathbb{C}[y]/{\frac{d Q}{d y}}.
\ee
The corresponding manifold $M$ is called the Frobenius manifold of $A_{q-1}$ type
\edefe 
The polynomials
\be
P_\alpha(y)=  \frac{\partial Q}{\partial u_\alpha },
\ee
form a basis in the tangent space $T_{\bf v} M$.  An invariant bilinear form (which is equivalent to the metric) 
is defined by
\be
(P_\alpha,P_\beta)=\underset{y=\infty}{\text{res}}\bigg(\frac{P_\alpha(y)P_\beta(y)}{\frac{dQ}{dy}(y)}\bigg).
\ee
With this definition one can verify that the corresponding metric is flat and
\be 
C_{\alpha\beta\gamma}=\nabla_\alpha\nabla_\beta\nabla_\gamma F(u).
\ee
To this end we perform the transformation from the canonical coordinates $ \{u_\alpha\} $ to the new coordinates $\{v^\alpha\}$ 
by means of the following relation
\begin{equation}\label{transform}
y=z -
\frac{1}{q}\bigg(\frac{v^{q-1}}{z}+\frac{v^{q-2}}{z^2}+\dots
+\frac{v^{1}}{z^{q-1}}\bigg)
+\mathcal{O}\bigg(\frac{1}{z^{q+1}}\bigg),
\end{equation}
where $ z^q=Q(y)$.

Some useful properties of the new coordinates are formulated in the following 
\th{properties} From the transformation \eqref{transform} it follows that
\begin{enumerate}
\item
 $v^\alpha$ form flat coordinates, i.e., the metric in this coordinates is constant and
\be
\eta_{\alpha\beta}=-q\bigg(\frac{\partial Q}{\partial v^\alpha},\frac{\partial Q}{\partial v^\beta}\bigg)=\delta_{\alpha+\beta,q},
\ee
\item 
\be
C_{\alpha\beta\gamma}=-q\underset{y=\infty}{\text{res}}
\bigg(\frac{\frac{\partial Q}{\partial v^\alpha} \frac{\partial Q}{\partial v^\beta}
\frac{\partial Q}{\partial v^\gamma}}{\frac{d Q}{dy}}\bigg)=
\frac{\partial^3 F}{\partial v^\alpha\partial v^\beta\partial v^\gamma}.
\ee
\item
\begin{equation}
\theta_{\alpha,k}=-c_{\alpha,k} \underset{y=\infty}{\text{res}} Q^{k+\frac{\alpha}{q}}(y),
\label{theta}
\end{equation}
where
\be
c_{\alpha,k}=\frac{\Gamma(\frac{\alpha}{q})}{\Gamma(\frac{\alpha}{q}+k+1)}.
\ee
\end{enumerate}
\eth
To prove these statements it is convenient to use the basis elements of $A_{q-1}$
in flat coordinates defined by $ \Phi_\alpha(y)= \frac{\partial Q(y)}{\partial v^\alpha}$  
which possess the following property 
\be
\Phi_\alpha(y)=\frac{1}{\alpha} \frac{d}{dy} \bigg(Q^{\frac{\alpha}{q}}\bigg)_{\!\!+}.
\ee

In what follows we use the following convention 
\be\label{conven}
\theta_{\mu,k} = \theta_{\mu -q \lfloor\mu/q\rfloor ,k+\lfloor\mu/q\rfloor},
\ee
where $\lfloor\mu/q\rfloor$ is the integer part of  $\mu/q$. 
It is clear that \eqref{conven} agrees with the definition  \eqref{theta}.

\section{Frobenius manifolds and Douglas string equation}
\label{sec:FM_IH}
\subsection{Integrable hierarchies} 
Let $\mathcal{M}$  be a space of functions of $x$ taking values in $M$.
 Let $I$ and $J$ be functionals on $\mathcal {M}$.
We define  the Poisson bracket  on $\mathcal {M}$ as
\be
\{I,J\}=\int \frac{\delta I}{\delta v^{\alpha}(x)} \eta^{\alpha\beta} \frac{d}{dx}  \frac{\delta I}{\delta v^{\beta}(x)} dx,
\ee
or 
\be
\{v^\alpha(x),v^\beta(y)\}=\eta^{\alpha\beta} \delta'(x-y),
\ee
where, as usual in the calculus of variations, the integrand is defined modulo  total derivatives.
 The functionals
\be
H_{\alpha,k}=\int \theta_{\alpha,k+1}(\vec{v}(x))dx, \qquad \alpha=1,...,n,\quad k\geq 0,
\ee
mutually commute among themselves
\be
\{H_{\alpha,k},H_{\beta,l}\}=0.
\ee
As a result, the Hamiltonian flows
\begin{eqnarray}\label{flows}
&\frac{\partial v^\mu}{\partial t^\alpha_k}=\{v^\mu,H_{\alpha,k}\}=\eta^{\mu\nu}\frac{\partial}{\partial x}
\frac{\partial \theta_{\alpha,k+1}}{\partial v^\nu}=C_\lambda^{\mu\rho}\frac{\partial \theta_{\alpha,k}}{\partial v^\rho}\frac{\partial v^\lambda}{\partial x}.
\end{eqnarray}
commute, i.e.,
\be\label{commflows}
\frac{\partial}{\partial t^\beta_l}\frac{\partial \vec{v}}{\partial t^\alpha_k}=
\frac{\partial}{\partial t^\alpha_k}\frac{\partial \vec{v}}{\partial t^\beta_l}.
\ee
It follows from \eqref{flows} that $t^1_0=x$.

\subsection{Douglas String Equation}
\label{sec:StrEqv}

Let us define a function $S(v,t)$  on $M$ which depends on the additional papametres   $\{t_k^\alpha\}$

\be
S(v,t^\alpha_k)=\sum_{\alpha=1}^n\sum_{k\geq0} t_k^\alpha \theta_{\alpha,k}(v).
\ee
The equation 
\be\label{streq}
\frac{\partial S}{\partial v^\alpha}=0,
\ee
is  called a string equation. 
In the case of Frobenius manifold of $ A_{q-1}$ type it is nothing but the Douglas string equation
  written in the form of the principle of least String action\cite{Ginsparg:1990zc}. 
It can be shown that solutions $\vec{v}(t^\alpha_k)$ of the string equation \eqref{streq}   satisfy also \eqref{flows}.

\subsection{Equation for Tau-function}
\label{sec:TauFun}

We define the function $Z[t]=\log \tau(t)$, where
\be
Z[t]=\frac{1}{2} \int_0^{v=v^*(t)} \Omega,
\ee
and
\be
\Omega=C_\alpha^{\beta\gamma}(v) \frac{\partial S(v,t)}{\partial v^\beta}
\frac{\partial S(v,t)}{\partial v^\gamma}  d v^{\alpha},
\ee
is the differential form and $v^*(t)$ is one of the solutions of the string equation \eqref{streq}.
From the associativity  of the algebra $ A_{q-1}$  and the equations \eqref{recur}
it follows that $\Omega$  is closed one-form.
\lem{Z(t)} On the solution of the string 
equation\footnote{To simplify our expressions 
we write $v(t)$ instead of $ v^*(t)$ when it is clear from the context.} 
Z(t) satisfies
\be
\frac{\partial^2 Z(t)}{\partial t_k^\alpha\partial t_0^1}=\theta_{\alpha,k}(v(t)).
\ee
In particular, 
\be
v^\alpha(t)=\eta^{\alpha\beta} \frac{\partial^2 Z}{\partial t_0^\beta\partial t^1_0},
\ee
and for $v^{q-1}(t)=u_1(t)$ 
\be\label{Z2}
\frac{\partial^2 Z}{\partial x^2}=u_1(t).
\ee
\elem
{\it Proof.} Differentiating with respect to $t_k^\alpha$ and $t_0^1$ and taking into account the string equation,
we find
\be
\frac{\partial^2 Z }{\partial t_k^\alpha\partial t_0^1}=
\int_0^{v^*(t)} C_{\lambda}^{\beta\gamma}
\frac{\partial \theta_{\alpha,k}}{\partial v^\beta} \frac{\partial \theta_{1,0}}{\partial v^\gamma}d v^\lambda=
\int_0^{v^*(t)}\frac{\partial\theta_{\alpha,k}}{\partial v^\lambda} d v^\lambda=\theta_{\alpha,k}.
\ee
Here we used that $\theta_{1,0}=v_1=v^{q-1}$, $C_\lambda^{\beta,q-1}=\delta_\lambda^\beta$.    $\square$

Taking into account $\theta_{\alpha,0}=v_{\alpha}$  we obtain from~\refl{Z(t)} that
\be
\frac{\partial^2 Z(t)}{\partial t_0^\alpha\partial t_0^1}=v_{\alpha}(t).
\ee
Since $Z$ satisfy equations \eqref{streq} and \eqref{Z2}, it is  a tau-function of the integrable hierarchy connected 
with the corresponding   Frobenius manifold.

\subsection{$A_{q-1}$ FM and  dispersionless limit of Gelfand-Dikij Hierarchy}
\label{sec:GD}

The dispersionless limit of the Gelfand-Dikij  equations is formulated as follows:
\be 
\frac{\partial Q}{\partial t_k^\alpha}=[A_{\alpha,k},Q]=\frac{\partial A_{\alpha,k}}{\partial x}\frac{\partial Q}{\partial y}
-\frac{\partial A_{\alpha,k}}{\partial y}\frac{\partial Q}{\partial x},
\ee
where
\be\label{Qpolynom}
Q=y^{q}+u_1(x)y^{q-2}+...+u_{q-1},
\ee
and
\be
A_{\alpha,k}=\frac{1}{q} c_{\alpha,k}\bigg(Q^{k+\frac{\alpha}{q}}\bigg)_+.
\ee
One can show that these equations are equivalent to the Hamiltonian equation
\be
\frac{\partial v^\mu}{\partial t_k^\alpha}=\eta^{\mu\nu} \frac{\partial}{\partial x} 
\frac{\partial \theta_{\alpha,k+1}}{\partial v^\nu}.
\ee

\subsection{Formula for tau-function}
\label{sec:FormulaTauFun}

As it was derived above the logarithm of the tau-function $Z[\{t^\alpha_k\}]$ is given by
\be
Z[\{t^\alpha_k\}]=\frac{1}{2}\int_0^{v^*(t)} C_{\alpha}^{\beta\gamma}\frac{\partial S}{\partial v^\beta}
\frac{\partial S}{\partial v^\gamma}d v^\alpha,
\ee
where
\be
S=\sum_{\alpha=1}^{q-1}\sum_k  t_k^\alpha \theta_{\alpha,k}\,.
\ee

\section{Resonance problem in $(p,q)$ MLG }
\label{sec:resonance}
\subsection{Homogeneity property of string equation. Spectrum for $(p,q)$ case  }

Let now only the finite number of the parameters $\{t_k^\alpha\}$ be nonzero. One of them we take equal to one and
others be enumerated by two integers $(m,n)$.  Here $1\leq m \leq q-1$,  $1\leq n\leq p-1$, 
where $p, q$ are two coprime integers,  $p>q$ and $q$ is a degree of the polynomial $Q$ defined in \eqref{Qpolynom}.  Hence,
the set of the parameters $\{t_k^\alpha\}$ is replaced by the set $\{t_{mn}\}$.
Let us take the action in the form
\be
S=\underset{y=\infty}{\text{res}}[Q^\frac{p+q}{q}+\sum_{m,n}^{pm-qn>0} t_{mn} Q^{\frac{pm-qn}{q}}],
\ee

It is easy to check that  $  Q[y,u_\alpha] $ and $ S[u_\alpha,t_{mn}] $ are  quasi-homogeneous
functions 
 \be
Q[\rho y,\rho^{r_\alpha} u_\alpha]=\rho^q Q[y,u_\alpha], \qquad  
S[\rho^{r_\alpha} u_\alpha,\rho^{\sigma_{mn}} t_{mn}]=\rho^{p+q} S[u_\alpha,t_{mn}].
\ee
Here we denote 
\be
r_\alpha=q-\alpha-1,\qquad \sigma_{mn}=p+q-|p m-q n|.
\ee
We call $\{\sigma_{mn}\}$ the set of the scaling indices of the set $\{t_{mn}\}$. As it was found by Douglas \cite{Douglas:1989dd},
the numbers $\delta_{mn}=\frac{\sigma_{mn}}{2q}$ coincide   with the gravitational dimensions of
the physical fields in $(p,q)$ Minimal Liouvillle gravity \cite {Knizhnik:1988ak}.

The function $Z[t_{mn}]$ is a quasi-homogeneous function 
\be
Z[\rho^{2q \delta_{mn}} t_{mn}]=\rho^{2(p+q)} Z[t_{mn}].
\ee

\subsection{ The group of the resonance transformations}
\label{sec:ResTrans}
Since the scaling indices are integer, the following relation can take place 
\be\label{resonances}
\sigma_{mn}=\sigma_{k_1l_1}+\sigma_{k_2l_2}+...+\sigma_{k_Nl_N}.
\ee
This is known as a resonance condition. The number of possible resonances
grows when $p$ and $q$ increase. 
A transformation $t_{mn}\rightarrow {\lambda}_{mn}$ of the form
\be
t_{mn}=\lambda_{mn}+\sum_{k_1,l_1,k_2,l_2} A_{mn}^{k_1l_1;k_2,l_2} 
\lambda_{k_1,l_1}\lambda_{k_2,l_2}+
\sum_{k_1,l_1,k_2,l_2,k_3,l_3} A_{mn}^{k_1l_1;k_2,l_2;k_3,l_3} \lambda_{k_1,l_1}\lambda_{k_2,l_2}\lambda_{k_3,l_3}+...,
\ee
is called resonance transformation  if \eqref{resonances} is satisfied for each term.
Besides, by definition, we  suggest that  the scaling index of  $ \lambda_{mn}$ equals to the one of $t_{mn}$.

It is obvious that
\be
t_{mn}(\{\rho^{\sigma_{kl}}\lambda_{kl}\})=
\rho^{\sigma_{mn}}t_{mn}(\{\lambda_{kl}\}),
\ee
and that the resonance transformation does not change the homogeneity property of the partition function
${Z}[t_{mn}(\{\lambda_{kl}\})]=\widetilde {Z}[\lambda_{mn}]$
\be
\widetilde{Z}[\{\rho^{\sigma_{mn}}\lambda_{mn}\}]=
\rho^{p+q}\widetilde{Z}[\{{\lambda}_{mn}\}].
\ee

Hence, if we find some solution of the string equation $ \eqref{streq}$ and 
construct $Z[t_{mn}]$, then we get a family of the solutions 
$\widetilde{Z}[\{\lambda_{mn}\}]=Z[\{t_{mn}(\{\lambda_{kl})\}\}]$
having the same homogeneity  properties with respect to the resonance transformations.

\subsection{The choice of the resonance transformation and of the solution of the string equation}
\label{sec:Choice}
Now we are in the position to formulate the following problem. We are looking for solutions of the string equation
and resonance transformations which gives function $\widetilde{Z}[\{\lambda_{mn}\}]$
satisfying infinite number of  constraints known as fusion rules for observables of minimal CFT models $M(p,q)$
and their for the correlators. In what follows we restrict ourself by considering spherical topology.
Then these rules can be formulated as follows.

We denote by $\Phi_{mn}$, where $1\leq m\leq p$ and $1\leq n\leq q$, the primary fields in the minimal
model $M(p,q)$ of Conformal field theory.
The fields $\Phi_{m,n}$ and $\Phi_{q-m,p-n}$ correspond to the same primary field. 

The following graphical representation allows to formulate these restrictions
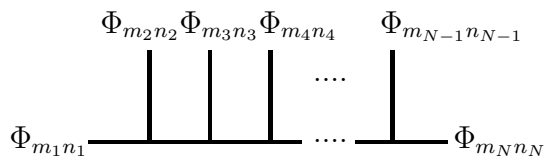
\begin{equation}\nonumber\label{conformal-block}
    \begin{picture}(-40,55)(80,10)
    \unitlength 2.3pt
    \linethickness{0.5mm}
    %%%%%%%%%%%%%%%%%%%%%%%
    \put(0,0){\line(1,0){35}}
       \put(44,0){\line(1,0){15}}
    %%%%%%%%%%%%%%%%%%%%%%%
    \put(10,0){\line(0,1){15}}
    \put(20,0){\line(0,1){15}}
    \put(30,0){\line(0,1){15}}   
    \put(50,0){\line(0,1){15}}   
    %%%%%%%%%%%%%%%%%%%%%%%
    \put(-13,-1){\mbox{$\Phi_{m_1n_1}$}}
       \put(60,-1){\mbox{$\Phi_{m_Nn_N}$}}
    \put(2,18){\mbox{$\Phi_{m_2n_2}$}}
    \put(15,18){\mbox{$\Phi_{m_3n_3}$}}
    \put(28,18){\mbox{$\Phi_{m_4n_4}$}}
    \put(48,18){\mbox{$\Phi_{m_{N-1}n_{N-1}}$}}
    %%%%%%%%%%%%%%%%%%%%%%%
   \put(37,10){\mbox{$....$}}
    \put(37,0){\mbox{$....$}}
    %%%%%%%%%%%%%%%%%%%%%%%%
%    \put(-40,10){$\mathcal{B}(\Delta_i;x)\equiv$}
    \end{picture}
    \vspace*{1cm}
\end{equation}
Here the external lines represent the (arbitrary arranged) primary fields in the correlator
$\langle \Phi_{m_1n_1}\Phi_{m_2n_2}...\Phi_{m_Nn_N}\rangle$  (here we assume $N\geq3$).
The fusion rules result to the requirement that the  correlation function  must be equal to zero 
if there are no sets of pairs $(k_i,l_i)$ assigned to the internal lines, for which in any
vertex of the graph the following condition on the three pairs $(m_i,n_i)$  ($i=1,2,3$) corresponding to the lines connected to this vertex 
\begin{align}
&|m_1-m_2|+1\leq m_3\leq \min\{m_1+m_2-1,2q+1-m_1-m_2\} \quad \text{step 2},\\
&|n_1-n_2|+1\leq n_3\leq \min\{n_1+n_2-1,2p+1-n_1-n_2\} \quad \text{step 2},
\end{align}
can  not be satisfied any permutation of the pairs.

In addition, from the conformal selection rules for $N=1$  it follows
\be
\langle \Phi_{mn}\rangle=0,
\ee
for $(m,n)\neq(1,1)$ and for $N=2$
\be 
\langle \Phi_{m_1n_1}\Phi_{m_2n_2}\rangle=0,
\ee
for $(m_1,n_1)$ not equal to $(m_2,n_2)$ or $(q-m_2,p-n_2)$. 

Now we are going to give a more precise formulation of our main conjecture. 
\conj{main} There exist the solution of the string equation
and the choice of the resonance transformation described above, such that the function
\be
\widetilde{Z}[\{\lambda_{mn}\}]=\langle \exp \sum_{m,n} \lambda_{m,n} O_{m,n} \rangle=
\sum_{N=0}^\infty \sum_{m_i,n_i} \frac{\lambda_{m_1n_1}...\lambda_{m_Nn_N}}{N!}
\langle O_{m_1n_1}...O_{m_Nn_N}\rangle,
\ee
appeares the generating function of the correlators in the Minimal Liouville Gravity.
\econj
In particular, all correlators $\langle O_{m_1n_1}...O_{m_Nn_N}\rangle$ forbidden by the conformal fusion rules vanish.

\section{The plan of the solution of the problem}
\label{sec:SolutionPlan}

To solve the formulated above problem we write the action $S(v_\alpha,t_{mn})$
and the generating function $Z[\{t_{mn}\}]$ in terms of new variables $ \{\lambda_{mn}\}$ using 
the resonance  change of variables 
\begin{align}
t_{mn}\,=\,\,&\lambda_{mn}+ A_{mn} \mu^{\delta_{mn}} + \!\!\!\!\!\sum_{m_1,n_1}^{\delta_{m_1n_1}\leq
\delta_{mn}} \!\!\!\!A^{m_1n_1}_{mn}\mu^{\delta_{mn}-\delta_{m_1n_1}} \lambda_{m_1n_1}
+\nonumber\\
&+\!\!\!\!\sum_{m_1,n_1,m_2,n_2}^{\delta_{m_1n_1}+\delta_{m_2n_2}\leq
\delta_{mn}} \!\!\!\!\!\!A^{m_1n_1,m_2n_2}_{mn}\mu^{\delta_{mn}-
\delta_{m_1n_1}-\delta_{m_2n_2}} \lambda_{m_1n_1}\lambda_{m_2n_2}+\dots,
\label{coupling}
\end{align}
where $\mu=\lambda_{11}$ is called the cosmological constant in the continuum approach to MLG. 

After performing this transform the action takes the form
\begin{align}\label{actionlambda}
\widetilde{S}[v_\alpha,\{\lambda_{mn}\}]=&S^{(0)}(v_\alpha)+\sum_{m,n} \lambda_{mn}\, S^{(mn)}(v_\alpha)+\nonumber\\
&+\sum_{m_1,n_1,m_2,n_2} \lambda_{m_1n_1}\lambda_{m_2n_2}\, S^{(m_1n_1,m_2n_2)}(v_\alpha)+\ldots.
\end{align}
The information about the form of the resonance transformation is encoded in the coefficients of  $S^{(0)}$, $S^{(mn)}$,
etc.  From \eqref{theta} and (12.1) we find
\begin{equation}\label{S0}
S^{(0)}\,\,=\,\,\underset{y=\infty}{\text{res}}\bigg[Q^{\frac{p+q}{q}}+ \sum_{l=1}^{s}
A_{1l}\,\, \mu^{ \frac{l+1}{2} } Q^{\frac{p - q l}{q}}\bigg],
\end{equation}
where we introduced the new positive integer numbers s and $p_0$ such that $p = s q + p_0$ and $0< p_0 < q$.
\begin{equation}\label{Smn}
S^{(mn)}=\underset{y=\infty}{\text{res}}\bigg[Q^{\frac{p m-q n}{q}}+
 \sum_{l=n+2}^{s m + \lfloor \frac{p_0 m}{q}\rfloor} A_{ml}^{mn}\,\, \mu^{\frac{l-n}{2}} \,\, Q^{\frac{ p m-q l}{q}}\bigg],
\end{equation}
where $A_{kl}^{mn}$ are the coefficients of the resonance relations and $(l-n)$ is even. The higher coefficients can also be easily
written in terms of the coefficients $A_{kl}^{\{m_in_i\}}$. 

The generating function
is given by
\begin{equation}
\widetilde{Z}[\{\lambda_{mn}\}]=\frac 12 \int_0^{\bf{v}^*} C_{\alpha}^{\beta\gamma}(v) \frac{\partial\widetilde{S}}
{\partial v^{\beta}}
 \frac{\partial\widetilde{S}}{\partial v^{\gamma}} d v^{\alpha},
\label{Z}
\end{equation} 
where ${\bf{v}^*}$ is defined  as a function of the parameters $\{\lambda_{mn}\}$ of
the Douglas string equation \eqref{streq}.

From now on we will skip the tilde over the functions
$\widetilde{S}(\{u_\alpha\},\{\lambda_{mn}\})$ and  $\widetilde{Z}(\{\lambda_{mn}\})$.

\section{Appropriate solution}
\label{sec:AppSol}

To compute the one-point function which is given by the integral 
\begin{eqnarray}
\langle O_{mn} \rangle=\int_0^{v_\alpha^0} C^{\alpha}_{\beta\gamma} \frac{\partial S^{(0)}}{\partial v_{\beta}}
\frac{\partial S^{(mn)}}{\partial v_{\gamma}} d v_{\alpha},
\label{Omn}
\end{eqnarray}
we need to know  the upper limit  in this integral $ v_\alpha^0 $ which  is the solution of the string equation 
for all couplings (except $\lambda_{11}=\mu $) equal to zero
\be
v_\alpha^0=v_\alpha^*(\lambda_{mn})\bigg|_{\lambda_{mn}=0,\lambda_{11}=\mu}.
\ee
Explicitly, $ v_\alpha^0 $ satisfies 
\be\label{streq1}
\frac{\partial S^{(0)}}{\partial v_\mu}\bigg|_{v_\alpha=v_\alpha^0}=0.
\ee
Using \eqref{S0}, \eqref{Smn} and \eqref{theta},  $S^{(0)} $ and $ S^{(mn)} $ can be written  as
\begin{align}\label{S0new}
&S^{(0)}=
-\frac{ \theta_{p_0,s+1}}{c_{p_0,s+1}}-
\sum_{l=1}^s A_{1l}\mu^{\frac{l+1}{2}}\frac{\theta_{p_0,s-l}}{c_{p_0,s-l}},\\
&S^{(mn)}=
-\frac{ \theta_{p_0 m ,s m - n}}{c_{p_0 m,s m-n}} -
\sum_{l=n+2}^{s m +  \lfloor\frac{p_0 m}{q}\rfloor} A_{ml}^{mn}\mu^{\frac{l+1}{2}}\frac{\theta_{p_0 m ,s m -l}}{c_{p_0 m ,s m-l}}.
\end{align}
We will use the following proposition  from~\cite{VBelavin:2014fs}:
\prop{theta}
 On the line $v_{i>1}=0$,
\begin{eqnarray}
\begin{cases}
k\text{ even}:&
\frac{\partial \theta_{\lambda,k}}{\partial v_{\alpha}}=\delta_{\lambda,\alpha} \,
x_{\lambda,k}\,\big(-\frac{v_1}{q}\big)^{\frac{k}{2}q},
\\
k\text{ odd}:&
\frac{\partial \theta_{\lambda,k}}{\partial v_{\alpha}}=\delta_{\lambda,q-\alpha} \,
y_{\lambda,k}\,\big(-\frac{v_1}{q}\big)^{\frac{k-1}{2}q+\lambda},
\end{cases}
\label{Lemma3}
\end{eqnarray}
where
\begin{eqnarray}\label{xy}
x_{\alpha,k}=\frac{\Gamma(\frac{\alpha}{q})}{\Gamma(\frac{\alpha}{q}+\frac{k}{2}) \big(\frac{k}{2}\big)! }
\qquad\text{and}\qquad
y_{\lambda,k}=-\frac{\Gamma(\frac{\alpha}{q})}{\Gamma(\frac{\alpha}{q}+\frac{k+1}{2}) \big(\frac{k-1}{2}\big)! }.
\label{Lemma31}  
\end{eqnarray}
\eprop
Using this statement together with   \eqref{S0new}
it is not difficult to see that the string equation  \eqref{streq1} has  the   solutions of the form $v^0_\alpha=0$ for $\alpha\neq1$
and the coordinate  $v^0_1$ is a root of the equation 
\be\label{pS01}
\frac{\partial S^{(0)}}{\partial v_{p_0}}=0 ,\quad \text{if}  \qquad s - \text{odd},
\ee
or
\be\label{pS02}
\frac{\partial S^{(0)}}{\partial v_{q-p_0}}=0 ,\quad \text{if}  \qquad s - \text{even}.
\ee
Here we assume that after taking derivative  we  set all $v_\alpha $ for   $\alpha\neq1$ to zero.
More explicitly these equations can be written as
\be\label{B1}
\sum_{k=-1:2:s} B^{\text{odd}}_{p_0,k} \bigg(-\frac{v_1}{q}\bigg)^{\frac{s-k}{2}q}=0, 
\quad \text{if}  \qquad s - \text{odd},
\ee
or 
\be\label{B2}
\sum_{k=-1:2:s} B^{\text{odd}}_{p_0,k} \bigg(-\frac{v_1}{q}\bigg)^{\frac{s-k-1}{2}q}=0, 
\quad \text{if}  \qquad s - \text{even},
\ee
where
\be
B^{\text{odd}}_{p_0,k}=\frac{x_{p_0,s-k}}{c_{p_0,s-k}} A_{1,k} \mu^{\frac{k+1}{2}},
\ee
and
\be
B^{\text{even}}_{p_0,k}=\frac{y_{p_0,s-k}}{c_{p_0,s-k}} A_{1,k} \mu^{\frac{k+1}{2}},
\ee
where $A_{1,-1}=1$.

\section{One-point functions}

As it was shown in \cite{VBelavin:2014fs}, the structure constant in the flat coordinates
on the line $v_{\alpha>0}=0$, for $\alpha\geq\beta\geq\gamma$
\begin{eqnarray}\label{strconstflat}
&C_{\alpha\beta\gamma}= 
\big(\!\!-\frac{v_1}{q}\big)^{\frac{\alpha+\beta+\gamma-q-1}{2}} \\
&\text{ if} \quad
\frac{\alpha+\beta+\gamma-q-1}{2}\in\mathbb{N}_0\quad\text{ and} \quad\alpha+\beta-\gamma\in[1,q-1],
\quad\text{ otherwise 0},\nonumber
\end{eqnarray}
where $\mathbb{N}_0$ is the set of non-negative integers.
 Using 
 \eqref{Lemma3} we find 
 for $s$ odd and $(sm-n)$ even 
\begin{eqnarray}
\langle O_{mn}\rangle\label{Omn1}
=\int_0^{v_1^0}C_{q-1,p_0,p_0 m}\frac{\partial S^{(0)}}{\partial v_{p_0}}
\frac{\partial S^{(mn)}}{\partial v_{p_0 m}} d v_{1}.
\label{Zmn1}
\end{eqnarray}
Taking into account \eqref{strconstflat} we conclude that the correlation function is zero for $m\neq1$.
Hence, in this case from the selection rules we obtain
\begin{eqnarray}\label{Zmn_od}
\langle O_{1n}\rangle
=\int_0^{v_1^{0}}\big(-\frac{v_1}{q}\big)^{p_0-1}\frac{\partial S^{(0)}}{\partial v_{p_0}}
\frac{\partial S^{(1n)}}{\partial v_{p_0 }} d v_{1}=0.
\end{eqnarray}
For $s$ odd and $(sm-n)$ odd, 
\begin{eqnarray}
\langle O_{mn}\rangle\label{Omn1odd}
=\int_0^{v_1^0}C_{q-1,p_0,q-p_0 m}\frac{\partial S^{(0)}}{\partial v_{p_0}}
\frac{\partial S^{(mn)}}{\partial v_{q-p_0 m}} d v_{1},
\end{eqnarray}
and the structure constant here is not equal to zero only if $q-p_0 m=p_0$ as it follows from \eqref{strconstflat}.
Therefore the gravitational dimension
\be 
[\langle O_{mn}\rangle]=\frac{p+q}{q}-\delta_{mn}=\frac{sm-n}{2}+\frac{s+1}{2}+\frac{p_0m+p_0}{2q},
\ee
is integer, the correlation function is analytic and we shell not  consider it\cite{Belavin:2013}.

Similarly, for $s$ even and $(sm-n)$ even, we obtain the following consequence of the selection rules
\begin{eqnarray}\label{Zmn_ev}
\langle O_{1n}\rangle
=\int_0^{v_1^{0}}\big(-\frac{v_1}{q}\big)^{q-p_0-1}\frac{\partial S^{(0)}}{\partial v_{q-p_0}}
\frac{\partial S^{(1n)}}{\partial v_{q-p_0}} d v_{1}=0.
\end{eqnarray}
Finally, if  $s$ even and $(sm-n)$ odd, we find again that the expressions for the one point correlation functions
are analytic.

Simple analysis shows that the number of these equations is equal to 
the number of the coefficients arising in the first order in the resonance relation.
Hence the requirement of absence of the one point functions fixes uniquely unknown coefficients ${B _{p_0,k}} $ in the expressions  \eqref{B1} and  \eqref{B2}. 

Thus we arrive to the conclusion that the special solution of the string 
equation considered above ensure the requirements of the selection rules in agreement with the general prescription described in the previous section. 

We note  also that the variety of $(p,q)$ models of minimal Liouville Gravity is splitted in two subclasses
according to the condition that $\lfloor p/q \rfloor$ be either even or odd. In each case we find
distinct sets of requirements formulated above leading to zero valued one point functions.

\section{Two-point functions}
We are now going to consider the two-point function. From \eqref{Z} we find
\begin{equation}\label{Z12}
\langle O_{m_1n_1}O_{m_2n_2}\rangle= \sum_{\gamma=1}^{q-1}
\int_0^{v_1^0} d v_1\, \big(-\frac{v_1}{q}\big)^{\gamma-1} \,
\frac{\partial S^{(m_1n_1)}}{\partial v_{\gamma}}\, \frac{\partial S^{(m_2n_2)}}{\partial v_{\gamma}}.
\end{equation}
It follows from \eqref{Lemma3} that
$\frac{\partial S^{(mn)}}{\partial v^{\gamma}}\neq0$ if one of the following two conditions is satisfied
\begin{equation}
\begin{aligned}\label{SmnCond}
&1)\quad\gamma=m p_0\, \text{mod}\, q\quad\text{and}\quad  (sm-n) - \text{even},\\
&2)\quad \gamma=q- m p_0\, \text{mod}\, q\quad\text{and} \quad (sm-n) - \text{odd}. 
\end{aligned}
\end{equation}
Similarly to the consideration in the previous section
we find four cases where the two point function can be non-zero. In two cases: where the first pair $(m_1,n_1)$ satisfies first condition while the second pair $(m_2,n_2)$ is subject of the second condition and vice versa, we find the regular expression for the two point function. 
Thus, we are  left with the two options  where both pairs satisfy either the first or the second condition in \eqref{SmnCond}.

Explicitly, in the  case when  both  $(sm-n_1) $  and  $(sm-n_2) $  are $ \text{even} $ we get the following requirement 
\begin{equation}
\langle O_{mn_1}O_{mn_2}\rangle= 
\int_0^{v_1^0} d v_1\,\big(-\frac{v_1}{q}\big)^{m p_0 -1}\,
\frac{\partial S^{(mn_1)}}{\partial v_{m p_0}}\, \frac{\partial S^{(mn_2)}}{\partial v_{m p_0}}=0\quad
\text{if}\quad n_1\neq n_2.
\end{equation}
Making the substitution
\be\label{t}
t=2\bigg(\frac{v_1}{v_{1}^0}\bigg)^q-1,
\ee
and denoting 
\be
\frac{\partial S^{(m n )}}{\partial v_{m p_0}}=  L_{\frac{s m-n}{2}}(t),
\ee
we find the following consequence of the diagonality condition
\begin{equation}\label{Z12_2}
\langle O_{mn_1}O_{mn_2}\rangle= 
\int_{-1}^{1} d t\,(1+t)^{\frac{m p_0-q}{q}}\,  L_{\frac{s m-n_1}{2}}(t)  L_{\frac{s m-n_2}{2}}(t)=0\quad
\text{if}\quad n_1\neq n_2.
\end{equation}
Hence, the selection rules  for  the
two-point correlation numbers requires that the polynomials $   L_{\frac{s m-n}{2}}$
form an orthogonal set of Jacobi  polynomials
\be
 \frac{\partial S^{(m n )}}{\partial v_{m p_0}}=\frac{pm-qn}{q} P_{\frac{s m-n}{2}}^{(0,\frac{m p_0-q}{q})}(t), 
\qquad\text{for}\quad    (s m-n)-\text{  even}.
\ee
In the second case,  where  both  $(sm-n_1) $  and  $(sm-n_2) $  are odd,
we have
\begin{equation}
\langle O_{mn_1}O_{mn_2}\rangle= 
\int_0^{v_1^0} d v_1\,\big(-\frac{v_1}{q}\big)^{q-m p_0-1}\,
\frac{\partial S^{(mn_1)}}{\partial v_{q-m p_0}}\, \frac{\partial S^{(mn_2)}}{\partial v_{q-m p_0}}=0\quad
\text{if}\quad n_1\neq n_2.
\end{equation}
Denoting
\be
\frac{\partial S^{(m n )}}{\partial v_{q-m p_0}}=(1+t)^{\frac{m p_0}{q}} L_{\frac{s m-n-1}{2}}(t),
\ee
we find the following consequence of the diagonality condition
for the two-point correlation function in this case
\begin{equation}
\langle O_{mn_1}O_{mn_2}\rangle= 
\int_{-1}^{1} d t\,(1+t)^{\frac{m p_0}{q}}\, L_{\frac{s m-n_1-1}{2}}(t) L_{\frac{s m-n_2-1}{2}}(t)=0\quad
\text{if}\quad n_1\neq n_2.
\end{equation}
It means  that
\be
 \frac{\partial S^{(m n )}}{\partial v_{q-m p_0}}=
 \frac{ pm-qn}{q} (1+t)^{\frac{m p_0}{q}}P_{\frac{s m-n-1}{2}}^{(0,\frac{m p_0}{q})}(t)
 \qquad\text{for}\quad    (s m-n)-\text{  odd}. 
\ee
At last,  inserting  these explicit expressions  for the derivatives of $S^{(m n )}$ to
the equations \eqref{Zmn_od} and \eqref{Zmn_ev} we arrive to the condition
\begin{eqnarray}
\langle O_{1n}\rangle
=\int_{-1}^{1} (1+t)^{\frac{p_0-q}{q}}L_{\frac{s+1}{2}}(t)
P_{\frac{s -n}{2}}^{(0,\frac{p_0-q}{q})}(t) d t= 0,
\end{eqnarray}
 in the case where $s$ is odd and $n$  is odd and greater than $1$.
And
\begin{eqnarray}
\langle O_{1n}\rangle
=\int_{-1}^{1}(1+t)^{\frac{p_0}{q}}L_{\frac{s}{2}}(t)
P_{\frac{s -n-1}{2}}^{(0,\frac{p_0}{q})}(t) d t=0,
\end{eqnarray}
in case where $s$ is even and $n$  is odd and greater than $1$.
Here  we introduced the polynomial $L_n(t)$
\be 
\frac{\partial S^{(0)}}{\partial v_{p_0}}(t)=L_{\frac{s+1}{2}}(t),
\ee
for $s$ odd,
\be 
\frac{\partial S^{(0)}}{\partial v_{q-p_0}}(t)=(1+t)^{\frac{p_0}{q}}L_{\frac{s}{2}}(t).
\ee
for $s$ even.

Taking to the account these equations, the order of the polynomials $\frac{\partial S^{(0)}}{\partial v_{p_0}} $
and $\frac{\partial S^{(0)}}{\partial v_{q-p_0}} $
and  the string equations \eqref{pS01},  \eqref{pS02} we obtain the following explicit expressions  
\begin{eqnarray}
 \frac{\partial S^{(0)}}{\partial v_{p_0}} =
\frac{p+q}{q}
\bigg(P_{\frac{s +1}{2}}^{(0,\frac{p_0-q}{q})}(t)-P_{\frac{s -1}{2}}^{(0,\frac{p_0-q}{q})}(t)\bigg),
\end{eqnarray}
if $s$ is odd and
\begin{eqnarray}
 \frac{\partial S^{(0)}}{\partial v_{q-p_0}} =\frac{p+q}{q}(1+t)^\frac{p_0}{q}
\bigg(P_{\frac{s }{2}}^{(0,\frac{p_0}{q})}(t)-P_{\frac{s -2}{2}}^{(0,\frac{p_0}{q})}(t) \bigg),
\end{eqnarray}
if $s$ is even.

\section{Conclusions}
\label{sec:Concl}

In this paper we have described the relation between the approach to $(p,q)$ models 
of Minimal Liouville gravity based on the Douglas string equation, on one hand, and
the Frobenius manifolds of $A_{q-1}$ type on the other. 
As a result  of this relation the generating function of correlation numbers in MLG
is represented by the logarithm of the tau-function of the corresponding integrable hierarchy.
All necessary information is encoded in the solution of the Douglas string equation and in the
resonance relations between the parameters of the integrable hierarchy and the coupling
constants of MLG. 
Using this relation and some special properties of the flat coordinates on the
Frobenius manifold, we have found the appropriate solution of the Douglas string equation. 
This result generalizes analogues result found recently for Unitary models of
Minimal Liouville gravity \cite{VBelavin:2014fs}. We have shown that the appropriate solution
is consistent with the basic requirements of the conformal selection rules
arising on the levels of one- and two-point correlation functions. Namely, the number
of the parameters of the resonance transformations is exactly the number of the constraints
following from the selection rules. Resolving these constraints we have found explicit 
form of the resonance transformations in terms of Jacoby polynomials.  
It would be interesting to investigate if this matching persists for  multi-point
correlation functions when
the fusion rules of the underlying minimal models of CFT should be taken into account.
This analysis requires also knowing the explicit form of the structure constants of the Frobenius algebra
in the flat coordinates. We plan to study these questions in the near future. 
Another possible extension of our study is to consider different generalizations of the
Minimal Liouville Gravity in the context of the Douglas string equation approach and its
relations with different types of Frobenius manifolds. In particular, it would
be interesting to understand what kind of the Frobenius manifold is relevant for 
$W_N$ Minimal Liouville gravity.

\vspace{5mm}

\noindent \textbf{Acknowledgements.}
We thank Boris Dubrovin, Michael Lashkevich, Yaroslav Pugai  and  Grisha Tarnopolsky for useful discussions.
The research was performed under a grant funded  by Russian Science Foundation 
(project No. 14-12-01383).

\vspace{5mm}

%%%%%%%%%%%%%%%%%%%%%%%%%%%%%%%%%%%%%%%%%%%%%%%%%%%%%%%%%%%%%%%%%%%%%%%%
%%%%%%%%%%%%%%%%%%%%%%%%%%%%%%%%%%%%%%%%%%%%%%%%%%%%%%%%%%%%%%%%%%%%%%%%
%%%%%%%%%%%%%%%%%%%%%%%%%%%%%%%%%%%%%%%%%%%%%%%%%%%%%%%%%%%%%%%%%%%%%%%%

\providecommand{\href}[2]{#2}\begingroup\raggedright
\addtolength{\baselineskip}{-3pt} \addtolength{\parskip}{-1pt}

\end{document}